\RequirePackage{fix-cm}
\documentclass[twocolumn]{svjour3}          
\usepackage{graphicx}
\usepackage{subcaption}
\usepackage{epstopdf}
\usepackage{hyperref}
\usepackage{flushend}

\journalname{Neural Computing and Applications}

\def\makeheadbox{{%
\hbox to0pt{\vbox{\baselineskip=10dd\hrule\hbox
to\hsize{\vrule\kern3pt\vbox{\kern3pt
\hbox{\bfseries Neural Computing and Applications}
\hbox{This is a post-peer-review, pre-copyedit version of this article.}
\hbox{The final authenticated version is available online at: \href{https://doi.org/10.1007/s00521-018-3464-7}{https://doi.org/10.1007/s00521-018-3464-7}.}
\kern3pt}\hfil\kern3pt\vrule}\hrule}%
\hss}}}

\begin{document}

\title{Towards Robust Voice Pathology Detection}

\subtitle{Investigation of supervised deep learning, gradient boosting, and anomaly detection approaches across four databases}

\author{
	Pavol~Harar \and
	Zoltan~Galaz \and
	Jesus~B.~Alonso-Hernandez \and
	Jiri~Mekyska \and
	Radim~Burget \and
	Zdenek~Smekal
}

\institute{
	P. Harar \at
  Brno University of Technology \\
	Technicka 3082/12 \\
	61600, Brno, Czech Republic \\
  \email{\href{mailto:pavol.harar@vut.cz}{pavol.harar@vut.cz}}
}

\date{Received: 10 January 2018 / Accepted: 24 March 2018}

\maketitle

\begin{abstract}
Automatic objective non-invasive detection of pathological voice based on computerized analysis of acoustic signals can play an~important role in early diagnosis, progression tracking and even effective treatment of pathological voices. In search towards such a~robust voice pathology detection system we investigated 3~distinct classifiers within supervised learning and anomaly detection paradigms. We conducted a~set of experiments using a~variety of input data such as raw waveforms, spectrograms, mel-frequency cepstral coefficients (MFCC) and conventional acoustic (dysphonic) features (AF). In comparison with previously published works, this article is the first to utilize combination of 4~different databases comprising normophonic and pathological recordings of sustained phonation of the vowel /a/ unrestricted to a~subset of vocal pathologies. Furthermore, to our best knowledge, this article is the first to explore gradient boosted trees and deep learning for this application. The following best classification performances measured by F1 score on dedicated test set were achieved: XGBoost (0.733) using AF and MFCC, DenseNet (0.621) using MFCC, and Isolation Forest (0.610) using AF. Even though these results are of exploratory character, conducted experiments do show promising potential of gradient boosting and deep learning methods to robustly detect voice pathologies.

\end{abstract}

\section{Introduction}
\label{introduction}

Voice pathology can be caused by the presence of tissue infection, systemic changes, mechanical stress, surface irritation, tissue changes, neurological and muscular changes, and other factors~\cite{Titze1994}. Due to vocal pathology, the mobility, functionality and shape of the vocal folds are affected resulting into irregular vibrations and increased acoustic noise. Such a~voice sounds strained, harsh, weak, and breathy~\cite{Teager1980,Hillenbrand1996}, which significantly contributes to the overall poor voice quality~\cite{brabenec2017speech,mekyska2015robust}.

Up to this day, vocal pathology detection has been approached by subjective and objective evaluations~\cite{mehta2008voice}. The first category (subjective evaluation) consists of so called in-hospital auditory-perceptual and visual examination of the vocal folds~\cite{oates2009auditory,song2013assessment}. For the visual examination laryngostroboscopy is commonly used~\cite{Uloza2015}. For the auditory-perceptual examination several clinical rating scales to diagnose and rate severity of vocal pathologies have been developed~\cite{de1997test,Gerratt1993,Kreiman1993,de1997test,Dejonckere2001}. Methods for subjective evaluation are, however, subject to inter-rater variability~\cite{armstrong1997place,gwet2014handbook}. Furthermore, they require patients' presence at the clinic, which can be a~serious problem especially in more severe stages of a~disease. This type of evaluation is also time costly, and it requires careful evaluation and scoring by clinicians.

\begin{table*}[htb!]
	\caption{Overview of related works focused on voice pathology detection.}
	\label{tab:related_Work}
	\scriptsize
	
	\begin{tabular}{p{2.1cm} l l l l l l}
		
		\hline\noalign{\smallskip}
		First author & Year & Ref. & Database & Input vowels & Classifier & Accuracy [\%] \\
		\hline\noalign{\smallskip}
		
			Hemmerling & 2016 & \cite{Hemmerling2016} & SVD & /a, i, u/ & KM, RF & 100.00 \\
			Muhammad & 2017 & \cite{Muhammad2017B} & SVD & /a/ & GMM & 99.98 \\
			Al-nasheri & 2017 & \cite{Alnasheri2017A} & MEEI, SVD, AVPD & /a/ & SVM & 99.81\,(MEEI), 91.17\,(AVPD), 90.98\,(SVD) \\
			Al-nasheri & 2017 & \cite{AlNasheri2017} & MEEI, SVD, AVPD & /a/ & SVM & 99.79\,(AVPD), 99.69\,(MEEI), 92.79\,(SVD) \\
			Al-nasheri & 2017 & \cite{Alnasheri2017B} & MEEI, SVD, AVPD & /a/ & SVM & 99.68\,(SVD), 88.21\,(MEEI), 72.53\,(AVPD) \\
			Eskidere & 2015 & \cite{Eskidere2015} & SVD & /a, i, u/ & GMM & 99.00 \\
			Amami & 2017 & \cite{amami2017incremental} & MEEI & /a/ & SVM & 98.00 \\
			Sabir & 2017 & \cite{sabir2017improved} & SVD & /a/ & ANN & 97.90 \\
			Hossain & 2016 & \cite{Hossain2016} & MEEI, SVD & /a, i, o/ & SVM, ELM, GMM & 95.00 \\
			Ali & 2017 & \cite{ali2017intra} & MEEI, SVD, AVPD & /a/ & GMM & 94.60\,(MEEI), 83.65\,(AVPD), 80.20\,(SVD) \\
			Muhammad & 2017 & \cite{Muhammad2017A} & MEEI, SVD, AVPD & /a/ & SVM & 99.40\,(MEEI), 93.20\,(SVD), 91.50\,(AVPD) \\
			Dahmani & 2017 & \cite{dahmani2017vocal} & SVD & /a/ & NB & 90.00 \\
			Souissi & 2016 & \cite{Souissi2016} & SVD & /a/ & ANN & 87.82 \\
			Hemmerling & 2017 & \cite{hemmerling2017voice} & SVD & /a/ & ANN & 87.50 \\
			Souissi & 2015 & \cite{Souissi2015} & SVD & /a/ & SVM & 86.44 \\
			
		\hline
	\end{tabular}
	
\vspace{0.5em}	
Table notation: Ref.\,--\,reference, MEEI\,--\,Massachusetts Eye and Ear Infirmary Database~\cite{eye1994voice,mekyska2015robust}, SVD\,--\,Saarbruecken Voice Database~\cite{Woldert2007,Muhammad2017A,Alnasheri2017A}, AVPD\,--\,Arabic Voice Pathology Database~\cite{mesallam2017development,Muhammad2017A}, KM\,--\,K-means~\cite{hartigan1979algorithm}, RF\,--\,Random Forests~\cite{breiman2001random}, GMM\,--\,Gaussian Mixture Models~\cite{reynolds2015gaussian}, SVM\,--\,Support Vector Machines~\cite{hearst1998support}, NB\,--\,Naive Bayes~\cite{murphy2006naive}, ELM\,--\,Extreme Learning Machine~\cite{huang2006extreme}, and ANN\,--\,Artificial Neural Networks~\cite{schalkoff1997artificial}.
	
\end{table*}

The second category (objective evaluation) is based on the automatic non-invasive computerized analysis of acoustic signals to quantify and identify the underlying vocal pathology that may not even be audible to a~human being~\cite{mekyska2015robust}. This type of evaluation is therefore inherently free from the subjective bias. Moreover, voice can be nowadays easily recorded using a~variety of smart devices, and processed remotely using cloud technologies. From these reasons, works such as~\cite{Eskidere2015,Hemmerling2016,Muhammad2017A,Alnasheri2017A} focused on using signal processing techniques (to quantify vocal-manifestations of the pathology under focus) and machine learning algorithms (to automate the process of voice pathology detection) to build a~system capable of accurate discrimination of healthy and pathological voices. In Table~\ref{tab:related_Work}, we summarize recent (2015\,--\,now) related works focused on the objective voice pathology detection.

For the purpose of the objective voice pathology evaluation, several databases have been frequently used by the researchers. Massachusetts Eye and Ear Infirmary Database (MEEI)~\cite{eye1994voice,mekyska2015robust}, Saarbruecken Voice Database (SVD)~\cite{Woldert2007,Muhammad2017A,Alnasheri2017A}, and Arabic Voice Pathology Database (AVPD)~\cite{mesallam2017development,Muhammad2017A} are among the most commonly used ones. More specifically, most researchers have analyzed sustained phonation of the vowel /a/, e.g.~\cite{Alnasheri2014,Muhammad2017B,amami2017incremental,dahmani2017vocal} due to its presence in most databases (language-independent speech task~\cite{Titze1994}). Some researchers also analyzed a~combination of the vowels, e.g.~\cite{Martinez2012,Eskidere2015,Hemmerling2016}, etc. From the voice pathologies point of view, most researchers restricted the dataset to a~limited set of pathologies~\cite{amami2017incremental,Muhammad2017B,dahmani2017vocal,hemmerling2017voice,sabir2017improved,Muhammad2017A,ali2017intra,AlNasheri2017,Alnasheri2017B,Alnasheri2017A}.

Next, conventional and clinically interpretable~\cite{brabenec2017speech}) acoustic features were usually computed prior to pathology detection~\cite{Muhammad2017B,dahmani2017vocal,sabir2017improved}. The acoustic features such as multidimensional voice program parameters (MDVP)~\cite{Alnasheri2017B}, mel-frequency cepstral coefficients (MFCC)~\cite{saldanha2014vocal}, glottal-to-noise excitation ratio (GNE)~\cite{michaelis1997glottal}, etc. were usually extracted. For more information about methods for pathological speech parametrization, see~\cite{mekyska2015robust}. After the feature extraction, multiple conventional classifiers have been used to detect the presence of voice pathology. Most authors relied on the following algorithms: Support Vector Machines (SVM), Gaussian Mixture Models (GMM), Random Forests (RF), and Artificial Neural Networks (ANN)~\cite{hemmerling2017voice,ali2017intra,amami2017incremental,dahmani2017vocal}, etc.

As can be seen, the results vary greatly between the published papers mainly due to differences in selected voice pathology samples, acoustic features, and classifiers that were used for the experiment. However, we can conclude that:
\begin{enumerate}
	\item most works analyzed a~single speech task, mainly the sustained phonation of the vowel /a/ (language independent speech task)
	\item most works analyzed datasets that were restricted to a~subset of vocal pathologies from 1~to 3~databases (MEEI, SVD, AVPD)
	\item most works extracted conventional dysphonic features to quantify major vocal-manifestations of specific vocal pathologies
	\item most works employed conventional supervised learning algorithms such as the following: SVM, GMM, RF, ANN, and others
\end{enumerate}

\begin{figure*}[tb!]
	\centering
	\begin{subfigure}[t]{0.45\textwidth}
			\includegraphics[width=\textwidth]{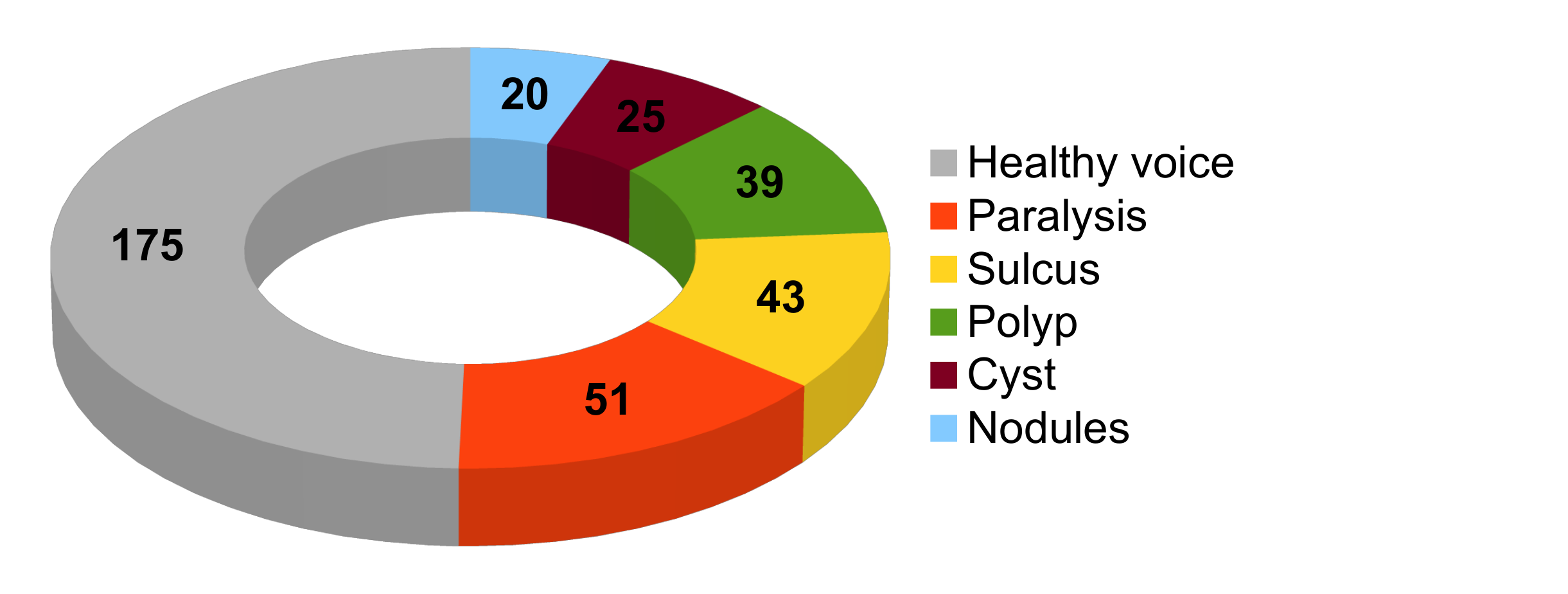}
			\caption{AVPD}
			\label{fig:avpd}
	\end{subfigure}
	\quad
	\begin{subfigure}[t]{0.45\textwidth}
			\includegraphics[width=\textwidth]{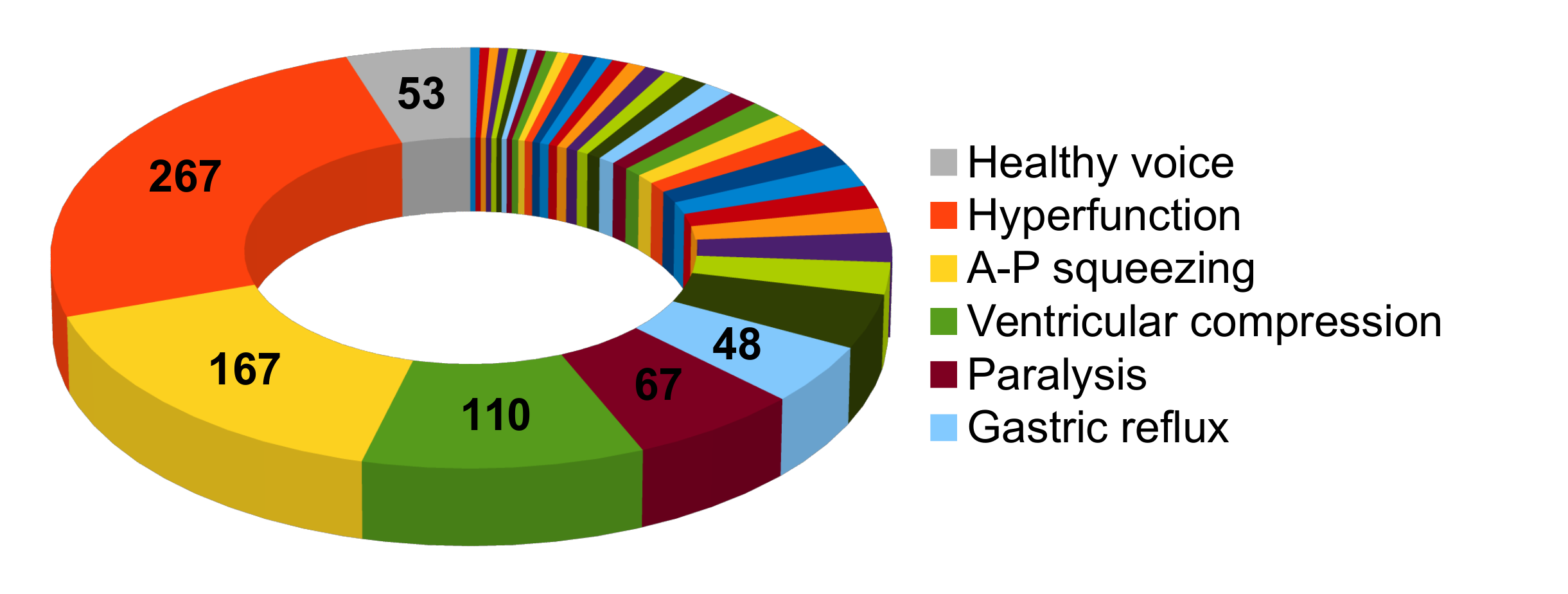}
			\caption{MEEI}
			\label{fig:meei}
	\end{subfigure}
		
	\begin{subfigure}[t]{0.45\textwidth}
			\includegraphics[width=\textwidth]{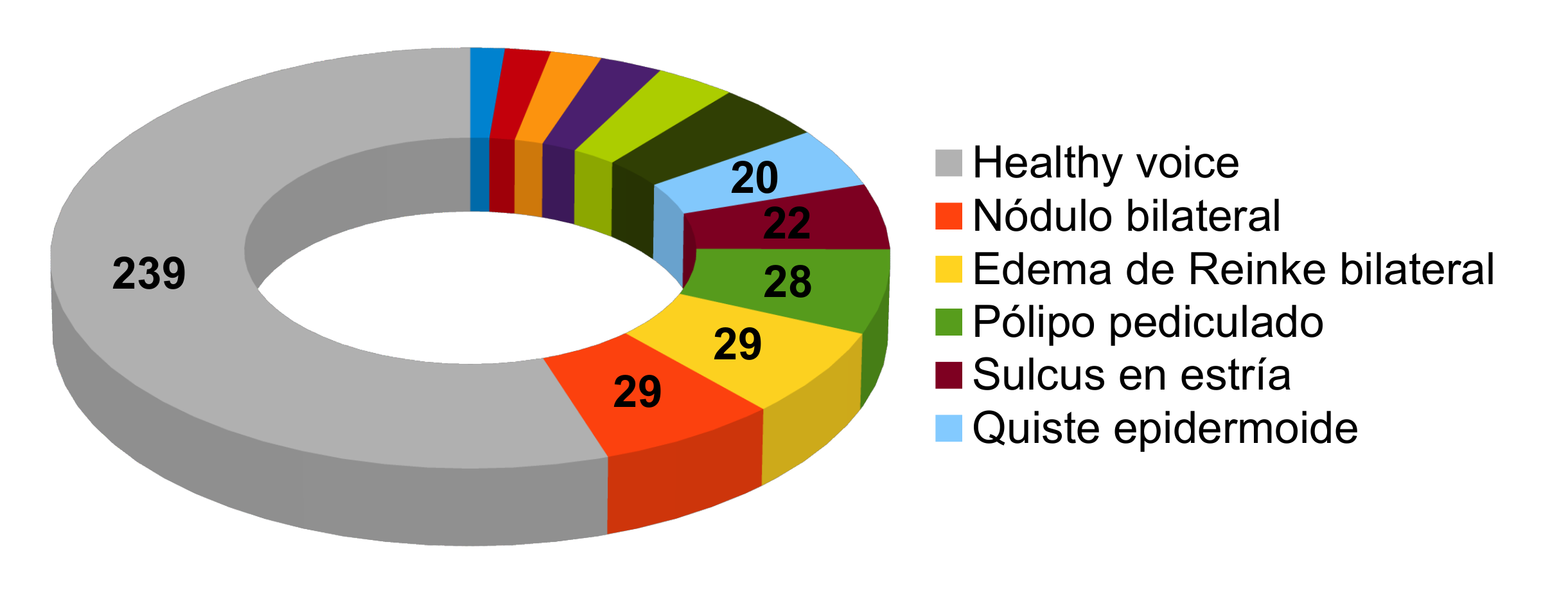}
			\caption{PDA}
			\label{fig:pda}
	\end{subfigure}
	\quad
	\begin{subfigure}[t]{0.45\textwidth}
			\includegraphics[width=\textwidth]{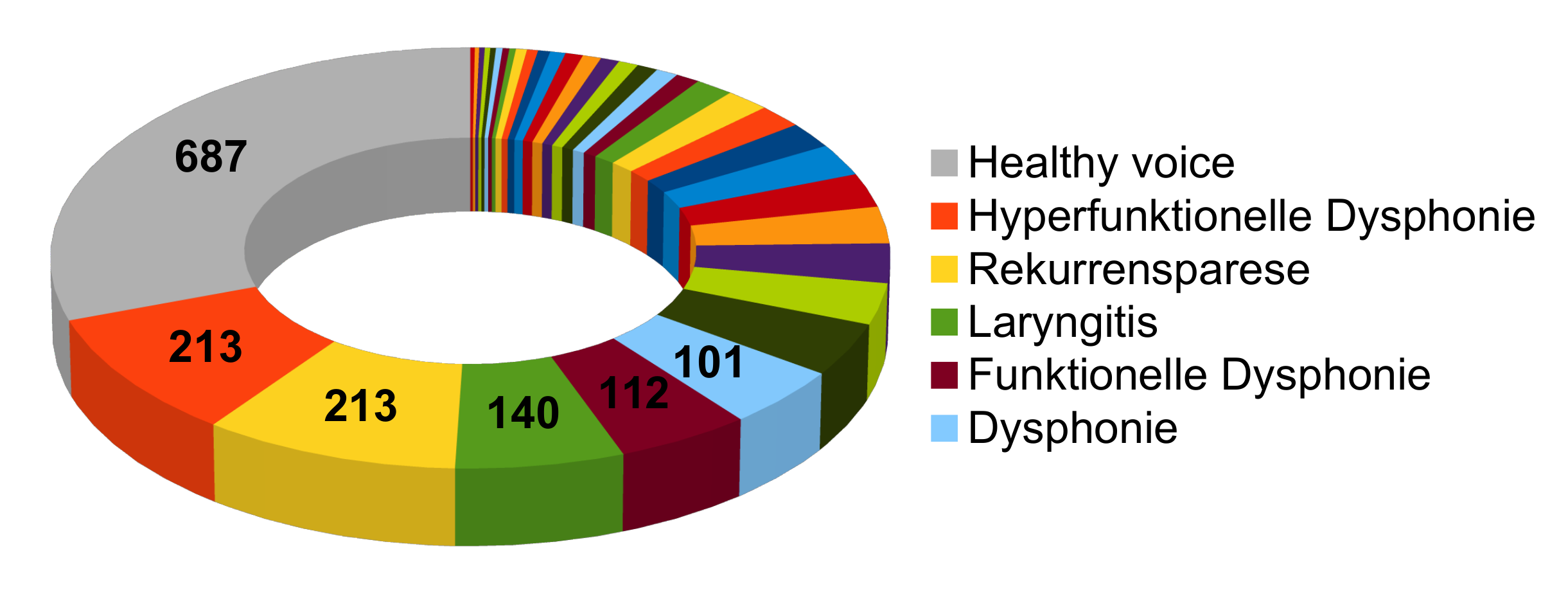}
			\caption{SVD}
			\label{fig:svd}
	\end{subfigure}
	
	\label{fig:databases}
	\caption{Visualization of inequality of samples per vocal pathology in the datasets used in this work (only 5~most common pathologies in each database are present in the legend), and healthy samples. Databases: a) AVPD~\cite{mesallam2017development,Muhammad2017A}, b) MEEI~\cite{eye1994voice,mekyska2015robust}, PDA~\cite{godino2010pathological,arias2011combining,mekyska2015robust}, and SVD~\cite{Woldert2007,Muhammad2017A,Alnasheri2017A}.}

\end{figure*}

To propose results comparable with the previously published works, we analyze voice recordings of sustained phonation of the vowel /a/ as well. However, unlike the previous works, we analyze a~larger dataset composed of 4~different databases, namely: MEEI~\cite{eye1994voice,mekyska2015robust}, SVD~\cite{Woldert2007,Muhammad2017A,Muhammad2017A,Alnasheri2017A}, AVPD~\cite{mesallam2017development,Muhammad2017A} (these databases are commonly used by the community), and Pr\'{i}ncipe de Asturias Database (PDA)~\cite{godino2010pathological,arias2011combining,mekyska2015robust}. Furthermore, to propose models capable of robust voice pathology detection, we do not restrict the dataset to only a~subset of common vocal pathologies. With this approach, our dataset does contain a~large number of pathologies with only few recordings. To see the sparsity of distribution and inequality of the number of pathologies in the databases, see Figures~\ref{fig:avpd} (AVPD),~\ref{fig:meei} (MEEI),~\ref{fig:pda} (PDA), and~\ref{fig:svd} (SVD).

By using 4~different databases, we aim to increase the size of our dataset to enable exploring possibilities of using supervised deep learning techniques that delivered state-of-the-art results in many domains including speech processing. To our best knowledge, despite our previous work~\cite{harar2017voice}, there are no other papers using deep learning algorithms for voice pathology detection. Next, we also employ the conventional voice pathology detection approach based on acoustic feature extraction procedure. However, unlike previous works, we use gradient boosting techniques for classification. To tackle the problem of sparse distribution of a~variety of vocal pathologies with only few recordings across the databases, we also investigate usage of anomaly detection procedure.

The rest of this paper is organized as follows. Section \ref{databases} introduces databases utilized in this article. In Section \ref{methodology}, the methodology of the experiment is discussed. The results are presented in Section \ref{results}. Conclusions are drawn in Section \ref{conclusions}.

\section{Databases}
\label{databases}

As mentioned previously, we chose the following speech task: sustained phonation of the vowel /a/ as a~basis for our experiments. During this particular speech task a~speaker is asked to sustain phonation of a~vowel, attempting to maintain steady frequency and amplitude at a~comfortable level~\cite{Titze1994}. The advantage of this speech task in comparison with other common speech tasks such as reading tasks, or a~running speech is that it is free of articulatory and other linguistic confounds~\cite{Titze1994}. This independence makes this task an ideal choice to construct a~large dataset that is necessary for supervised deep learning algorithms. In fact, sustained /a/ vowel phonation is the only speech task that is present in all databases used in this work. The contents of the databases relevant to this work can be seen in Table\,\ref{tab:contents}.

\subsection{AVPD database}
\label{avpd}

Arabic Voice Pathology Database (AVPD)~\cite{mesallam2017development,Muhammad2017A} was developed at the Communication and Swallowing Disorders Unit of King Abdul Aziz University Hospital, Riyadh, Saudi Arabia. The database contains recordings (366 samples: 188 healthy, 178 pathological) of sustained phonation of the vowels /a, e, o/, counting from 0-10, standardized Arabic passage, and reading three words. All recordings are sampled at $f_s = 48\,000$\,Hz with a bit depth of 16 bits. The database comprises five organic voice disorders: vocal fold cysts, nodules, paralysis, polyps, and sulcus. Multiple recordings of the same vowel were taken to help model the intra-speaker variability.

\subsection{MEEI database}
\label{meei}

Massachusetts Eye and Ear Infirmary Database (MEEI) \cite{eye1994voice,mekyska2015robust} is one of the most popular and most widely-used database (used for many years as a~benchmark in the field of pathological speech analysis). It contains more than 1\,400 recordings of sustained phonation of the vowel /a/ (recorded from 657 pathological speakers with different types of pathologies and 53 healthy speakers). This database has several disadvantages such as the fact that recordings of the normophonic voice were recorded in different conditions (e.g. different environment, recordings are sampled at: $f_s = 50\,000$\,Hz, $f_s = 25\,000$\,Hz, and $f_s = 10\,000$\,Hz) when compared to pathological recordings. The database is also gender-unbalanced, etc.

\subsection{PDA database}
\label{pda}

Pr\'{i}ncipe de Asturias Database (PDA)~\cite{godino2010pathological,arias2011combining,mekyska2015robust} contains recordings of 200 pathological speakers with different types of organic pathologies (e.g. nodules, polyps, oedemas, and carcinomas, etc.) and 239 healthy speakers. For each speaker, sustained phonation of the vowel /a/ is recorded. All recordings are sampled at $f_s = 25\,000$\,Hz. This database contains more speakers than a~balanced version of MEEI database that according to~\cite{Parsa2003} comprise only 173 recordings of pathological speakers.

\subsection{SVD database}
\label{svd}

Saarbruecken Voice Database (SVD)~\cite{Woldert2007,Muhammad2017A,Alnasheri2017A} is a~collection of voice recordings and electroglottography (EGG) signals from more than 2\,000 speakers. It contains recordings of 687 healthy persons (428 females and 259 males) and 1356 patients (727 females and 629 males) with one or more of the 71 different pathologies. One session contains the recordings of the following components: a) vowels /i, a, u/ produced at normal, high and low pitch; vowels /i, a, u/ with rising-falling pitch; and c) sentence ``Guten Morgen, wie geht es Ihnen?'' (``Good morning, how are you?''). All samples of the sustained vowels are between 1~to 3~seconds long, sampled at $f_s = 50\,000$\,Hz with 16-bit resolution~\cite{Woldert2007}. In contrary to MEEI database, all audio samples (healthy and pathological) in SVD were recorded in the same environment. 

\begin{table}[tb!]
	\caption{Contents of the databases used in this work.}
	\label{tab:contents}
	
	\begin{tabular}{p{2.4cm} r r r r}
		\hline\noalign{\smallskip}
		& AVPD & MEEI & PDA & SVD \\
		\noalign{\smallskip}\hline\noalign{\smallskip}
		
		H samples & 188 & 53 & 239 & 687 \\
		P samples & 178 & 657 & 200 & 1356 \\
		vowels & /a, e, o/ & /a/ & /a/ & /a, i, u/ \\
		running speech & yes & yes & no & yes \\
		EGG & no & no & no & yes \\
		GRBAS & yes & no & no & no \\
		\noalign{\smallskip}\hline\noalign{\smallskip}
		
	\end{tabular}
	
\vspace{0.5em}
Table notation: PDA\,--\,Pr\'{i}ncipe de Asturias Database (PDA)~\cite{godino2010pathological,arias2011combining,mekyska2015robust}, MEEI\,--\,Massachusetts Eye and Ear Infirmary Database~\cite{eye1994voice,mekyska2015robust}, SVD\,--\,Saarbruecken Voice Database~\cite{Woldert2007,Muhammad2017A,Alnasheri2017A}, AVPD\,--\,Arabic Voice Pathology Database~\cite{mesallam2017development,Muhammad2017A}, H\,--\,healthy, P\,--\,pathological, and GRBAS\,--\,Grade, Roughness, Breathiness, Asthenia, Strain scale~\cite{de1997test}.

\end{table}

\section{Methodology}
\label{methodology}

\subsection{Data processing}
\label{data_processing}

We used 720 recordings from AVPD, 709 recordings from MEEI, 422 from PDA and 2\,040 from SVD. We only excluded samples that were shorter than 0.750\,s in length (removed 319 recordings). We also excluded all recordings of speakers bellow the age of 19 and also above the age of 60 (it is known that the most significant changes of voice happen during adulthood until the voice matures at around age of 20 and remains relatively stable until around age of 60) \cite{stathopoulos2011changes}. After these restrictions, the final number of samples equaled to 2\,707.

Using \href{http://sox.sourceforge.net/}{SOX library} (version 14.4.2), we re-sampled each recording to $f_s = 16\,000$\,Hz. Then we trimmed each sample to exactly 0.750\,s in duration. If a~recording was below 0.950\,s in duration, we extracted only one sample from the middle of it. For longer recordings we trimmed each end by 0.100\,s and extracted as many 0.750\,s long chunks as possible without overlap with stride of 0.375\,s. Using this approach, the total number of 8\,042 chunks was obtained. Further details regarding the number of chunks used can be found in Table\,\ref{tab:nchunks}.

\begin{table}[tb!]
	\caption{Number of chunks used in the experiments.}
	\label{tab:nchunks}

	\begin{tabular}{p{1.8cm} r r r r r}
		\hline\noalign{\smallskip}
		Database & H (M) & P (M) & H (F) & P (F) & Total \\
		\noalign{\smallskip}\hline\noalign{\smallskip}
		
		AVPD & 625 & 509 & 872 & 804 & 2810 \\
		MEEI & 126 & 114 & 185 & 168 & 593 \\
		PDA & 1158 & 331 & 5 & 605 & 2099 \\
		SVD & 400	& 645	& 624	& 871 & 2540 \\
		
		\noalign{\smallskip}\hline\noalign{\smallskip}
		Total	& 2309 & 1599 & 1686 & 2448 & \textbf{8042} \\
		\noalign{\smallskip}\hline\noalign{\smallskip}
	\end{tabular}
	
\vspace{0.5em}	
Table notation: PDA\,--\,Pr\'{i}ncipe de Asturias Database (PDA)~\cite{godino2010pathological,arias2011combining,mekyska2015robust}, MEEI\,--\,Massachusetts Eye and Ear Infirmary Database~\cite{eye1994voice,mekyska2015robust}, SVD\,--\,Saarbruecken Voice Database~\cite{Woldert2007,Muhammad2017A,Alnasheri2017A}, AVPD\,--\,Arabic Voice Pathology Database~\cite{mesallam2017development,Muhammad2017A}, H\,--\,healthy, P\,--\,pathological, M\,--\,males, and F\,--\,females.

\end{table}

\subsection{Feature extraction}
\label{feature_extraction}

At first, we considered raw audio samples as an input data for the voice pathology detection model. Each file (the 0.750\,s chunk) was therefore inserted to the input of the neural network as a~vector of 12\,000 features (computed as: 0.750\,s $\cdot$ 16\,000\,Hz = 12\,000 features). Additionally, we normalized each sample using min-max scaling to a~range $\langle 0, 1 \rangle$.

Next, we extracted a~set of conventional commonly-used acoustic (dysphonic) features~\cite{brabenec2017speech,mekyska2015robust} using Neurological Disorder Analysis Tool (NDAT)~\cite{Smekal2015b,mekyska2015robust} written in MATLAB and developed at the Brno University of Technology. Specifically, we computed the following acoustic features: pitch, jitter, shimmer, harmonic-to-noise ratio, detrended fluctuation analysis parameters, glottis quotients (open, closed), glottal-to-noise excitation ratio, Teager-Kaiser energy operator, modulation energy, and normalized noise energy. We further applied the following statistical properties: mean, standard deviation, coefficient of variation, quartiles ($1^{\mathrm{st}}$, $2^{\mathrm{nd}}$, $3^{\mathrm{rd}}$), interquartile range, kurtosis, and skewness.

Moreover, we computed spectrograms using \href{https://matplotlib.org/}{Matplotlib} (version 2.1.2) library for Python. The computation setup: mode (power spectral density), no windowing, no overlap, and NFFT ($512$\,samples). Following Ali et al.~\cite{ali2017automatic}, we used data up to 1\,500\,Hz (1\,150 features). Furthermore, we normalized the values of this matrix with min-max scaling to a~range between 0~and 1 as well.

At last, we computed most commonly used perceptual~\cite{Smekal2015b} acoustic feature: MFCC using \href{https://github.com/jameslyons/python_speech_features/}{Python Speech Features library}. The computation setup: length of a~window function ($0.025$\,s), step size ($0.010$\,s), number of filters in the filter-bank ($26$), number of coefficients ($13$), and NFFT ($512$\,samples). With this approach, we obtained a~matrix consisting of 962 features (13\,coefficients $\times$ 74\,time steps). We also computed the mean and standard deviation of the 13 coefficients along the time axis, which resulted into additional 26 features per sample. Next, we scaled the MFCC feature matrix by min-max algorithm (means and standard deviations were computed before scaling). The statistical features were scaled separately to have 0~mean and unit variance before classification.

\subsection{Experiments}
\label{experiments}

As mentioned previously, there is a~wide range of pathologies present in the databases used in this work. For more information, see Figures~\ref{fig:avpd} (AVPD),~\ref{fig:meei} (MEEI),~\ref{fig:pda} (PDA), and~\ref{fig:svd} (SVD). Each database was labelled in different language and with different experts by different criteria. Therefore, it is almost impossible to combine these databases to obtain one consistent database of multiple pathologies with reasonable number of samples. Only feasible way of combining the samples seems to be the exhaustive manual pairing by an expert clinician, which is also rather difficult since lots of recordings are labelled with multiple pathologies. In order to conduct inter-database experiments, authors therefore usually pick a~smaller sub-sample of 2~to 5~pathology types that are relatively easier to pair. 

Next, most of these pathologies are very sparsely distributed across the databases. Searching for an ideal subset of acoustic features that would yield a~good classification performance for each voice pathology is therefore almost impossible. Furthermore, it is not well-known if these pathologies present in the databases have similar vocal-manifestations.

In contrast to the previous works, we aim to investigate possibilities of robust voice pathology detection using a~set of 4~almost unrestricted databases comprising a~large number of pathologies. From these reasons, we decided to conduct several experiments: a) supervised learning (assuming the pathologies have similar manifestations and therefore the number of samples per pathology type is irrelevant), b) anomaly detection (assuming the pathologies do not have similar manifestations and therefore the number of samples per pathology type cannot be neglected).

Regarding the supervised learning approach, we used the state-of-the-art gradient boosting algorithm: XGBoost~\cite{chen2016xgboost} (version 0.6) for its current successes in many Kaggle competitions, fast training and model interpretability. Additionally, we explored possibilities of deep learning approach for its ability to robustly model complex relationships when optimized using enough data. However, the equation for computing the sufficient size of training dataset has not been formally described yet. Generally established rule of thumb in machine learning community is to have more training samples than trainable parameters. For this reason, we used the DenseNet~\cite{huang2016densely} architecture, which succeeded in overcoming the problem of having too many trainable parameters by densely connecting the convolutional layers. We adjusted \href{https://github.com/tdeboissiere/DeepLearningImplementations/tree/master/DenseNet}{Thibault de Boissiere's Keras implementation of the DenseNet} (Keras framework~\cite{Chollet2015}, version 2.1.2), to process 1D signals treating raw audio as 1D vector. Spectrograms were processed as a~matrix using the frequency bins not as y~dimension, but rather as a~stack of channels in the same way the 3~RGB channels are stacked in an image \cite{wyse2017audio}. The MFCC were processed the same way as spectrograms. Since we are not able to say with 100\% certainty that healthy examples are not polluted by deviant samples, we decided to use anomaly detection in favor of novelty detection in which case it is important to model the non-deviant samples. In this case, we chose Isolation Forest~\cite{liu2008isolation,liu2012isolation} classifier implemented in scikit-learn library~\cite{scikit-learn} (version 0.19.1).

For the above mentioned experiments, we decided to analyze the performance of the voice pathology detection models using multiple types of input data: a)~raw audio samples to follow our previous work~\cite{harar2017voice} and further explore possibilities of robust voice pathology detection without manually-selected features (DenseNet), b)~conventional acoustic (dysphonic) features to follow the previously published works and quantify most common vocal pathologies (XGBoost, Isolation Forest), c)~spectrograms to achieve a~reasonable trade-off between dimensionality of the data and amount of information (DenseNet), and d)~MFCC to follow the previous works focusing on voice and speech modelling, and voice pathology detection (all models). 

\subsection{Training and validation}
\label{training_and_validation}

To train and validate the models, we split the data to training, validation and testing sets. On top of that, we generated 10-fold validation indices using training and validation sets, so we can use exactly the same sets of data for each experiment. The test set was left for final evaluation of the models. Next, we stratified the testing and validation sets by medical state (healthy\,--\,H, pathological\,--\,P), gender, age, and gender-age group. Since the longer recordings were split into multiple chunks, we had to prevent the samples with chunks in the test or validation sets from leaking into the training set. Such chunks were carefully removed from the set. All other chunks were used in the training set.

At this point there is an unequal distribution of samples within the training set. We reacted to this fact by computing sample weights that can be used during training as a~compensation measure for under-represented groups. The final sample weight is a~product of 3~partial weights. Each of the partial weights quantifies the ratio between subgroups within the selected group (e.g. the ratio between the number of normophonic and pathological samples). For this purpose, we introduced a~class weight $\alpha$, gender weight $\beta$, and gender-age group weight $\gamma$ resulting in final sample weight $\omega$ that is computed as $\omega = \alpha \cdot \beta \cdot \gamma$. Weight for a~particular sample that belongs to subgroup $\alpha_i$ within group $\alpha$, $\beta_i$ within group $\beta$, and $\gamma_i$ within group $\gamma$ can be computed as follows:

\begin{equation} \label{eq:1}
\omega_{\alpha_i, \beta_i, \gamma_i} = \frac{n_{\alpha_i}}{\textrm{max}(n_{\alpha})} \cdot \frac{n_{\beta_i}}{\textrm{max}(n_{\beta})} \cdot \frac{n_{\gamma_i}}{\textrm{max}(n_{\gamma})}
\end{equation}
where $n$ represents the total number of samples: $n_{\alpha_i}$ is the number of samples in a~particular class; $n_{\beta_i}$ is the number of samples in a~particular gender; and $n_{\gamma_i}$ is the number of samples in a~particular gender-age group. 

We used 30 to 100 iterations of randomized cross-validation search for hyper-parameter optimization for both XGBoost and Isolation Forest classifiers. The number of iterations did depend on the fitting time. More specifically, in the case of XGBoost, we were searching over the following hyper-parameters: number of estimators~$\langle 3, 300 \rangle$, learning rate~$\langle 0.006, 1 \rangle$, gamma~$\langle 10, 60 \rangle$, maximum depth $\langle 0, 9 \rangle$, minimum child weight $\langle 1, 3 \rangle$, sub-sample ratio $\langle 0.3, 1 \rangle$ and colsample bytree (sub-sample ratio of columns when constructing each tree) $\langle 0.1, 1 \rangle$. Regarding Isolation Forest we were searching over the following hyper-parameters: number of estimators~$\langle 6, 200 \rangle$, maximum samples $\langle 8, 64 \rangle$, contamination~$\langle 0.40, 0.76 \rangle$ and maximum features~$\langle 0.05, 1 \rangle$. As a~performance measure, we used F1 micro score as a~criteria of choosing the best hyper-parameters in the cross-validation setup. After the search for hyper-parameter, we re-fitted the models with the best hyper-parameters on the entire training set, and consequently evaluated on the testing set. The final results are presented in the form of confusion matrix (CM), and classification report (CR) tables. The formulas \ref{eq:2}, \ref{eq:3} and \ref{eq:4} describe the way of computing the precision, recall and F1 score (weighted average of the precision and recall) metrics presented in CR tables.

\begin{equation} \label{eq:2}
precision = \frac{tp}{tp + fp},
\end{equation}
where $tp$ denotes the number of correct predictions (observed class), and $fp$ determines the number of incorrect predictions (observed class). The precision is a ratio between the number of correct predictions of the observed class and the total number of predictions of the observed class.

\begin{equation} \label{eq:3}
recall = \frac{tp}{tp + fn},
\end{equation}
where $tp$ denotes the number of correct predictions (observed class), and $fn$ determines the number of incorrect predictions (opposing class). The recall is a ratio between the number of correct predictions of the observed class and the total number of samples in the observed class.

\begin{equation} \label{eq:4}
F1 = 2 \cdot \frac{precision \cdot recall}{precision + recall}
\end{equation}

\section{Results}
\label{results}

XGBoost~\cite{chen2016xgboost} trained (10-fold validation) with all features (consisting of 96 conventional dysphonic features and 26 MFCC coefficients) yielded an average F1 score of $0.922\,(\pm\,0.004)$ on the training set, and $0.829\,(\pm\,0.028)$ on the validation set. The final F1 score on the dedicated testing set was $0.733$. Performance details (classification matrix and classification report) can be found in Table~\ref{tab:xgbconf}, and Table~\ref{tab:xgbreport}. Based upon the performance on the development set (training and validation sets) the 50 iterations of randomized cross-validated search selected the following hyper-parameters: number of estimators ($294$), learning rate ($0.3$), gamma ($10$), max. depth ($3$), sub-sample ($0.5$), minimum child weight ($1$), colsample bytree ($1$). Details regarding the classification performance in relation to input data can be found in Table~\ref{tab:xgboostinputdata}.

\begin{table}[htb!]
\caption{Testing CM for XGBoost}
\label{tab:xgbconf}
\begin{tabular}{p{2.6cm} l l l}
\hline\noalign{\smallskip}
& true H & true P & total predicted \\
\noalign{\smallskip}\hline\noalign{\smallskip}		
predicted H & \textbf{82} & 26 & 108 \\
predicted P & 38 & \textbf{94} & 132 \\	
\noalign{\smallskip}\hline\noalign{\smallskip}
total true & 120 & 120 & accuracy: \textbf{0.733} \\
\noalign{\smallskip}\hline\noalign{\smallskip}
\end{tabular}
\end{table}

\begin{table}[htb!]
\caption{Testing CR for XGBoost}
\label{tab:xgbreport} 
\begin{tabular}{lllll}
\hline\noalign{\smallskip}
& precision & recall & f1-score & no. samples \\
\noalign{\smallskip}\hline\noalign{\smallskip}
class H & 0.759 & 0.683 & 0.719 & 120\\
class P & 0.712 & 0.783 & 0.746 & 120\\
\noalign{\smallskip}\hline\noalign{\smallskip}
avg. / total & 0.736 & 0.733 & \textbf{0.733} & 240 \\
\noalign{\smallskip}\hline
\end{tabular}
\end{table}

Regarding deep learning approach, we used the adjusted DenseNet~\cite{huang2016densely} architecture with the binary cross-entropy loss optimized using Adam optimizer \cite{Kingma2014}. The initial learning rate was set to $0.01$ with decay of $1\textrm{e}-04$ on each epoch. Hyper-parameter optimization was done using training and validation sets, and the final parameters of the DenseNet network were set as follows: depth ($4$), number of dense blocks ($2$), growth rate ($5$), number of filters ($10$), drop-out rate ($0.3$), l2~weight decay ($1\textrm{e}-04$). The input shape of this network was $(13 \times 47)$ with one neuron in the last layer with sigmoid activation function, and the total of $1\,629$ trainable parameters. For this particular setup with MFCC as the input data, the system yielded F1 score of $0.595$ on the training set, and $0.648$ on the validation set. After the hyper-parameter optimization, we retrained the network on all data from the training and validation sets (the development set), and the system yielded the final F1 score on the dedicated testing set of $0.621$. Performance details (classification matrix and classification report) can be found in Table\,\ref{tab:densenetconf} and Table\,\ref{tab:densenetreport}.

\begin{table}[tb!]
\caption{XGBoost performance related to input data}
\label{tab:xgboostinputdata} 
\begin{tabular}{llll}
\hline\noalign{\smallskip}
Input data & F1 CV train & F1 CV valid & F1 test \\
\noalign{\smallskip}\hline\noalign{\smallskip}
ALL 		& $0.922\,(\pm\,0.004)$ & $0.829\,(\pm\,0.028)$ & \textbf{$0.733$}\\
AF stats 	& $0.886\,(\pm\,0.004)$ & $0.791\,(\pm\,0.034)$ & $0.686$\\
AF			& $0.892\,(\pm\,0.006)$ & $0.798\,(\pm\,0.025)$ & $0.658$\\
AF base 	& $0.745\,(\pm\,0.009)$ & $0.689\,(\pm\,0.036)$ & $0.646$\\
MFCC		& $0.680\,(\pm\,0.010)$ & $0.769\,(\pm\,0.037)$ & $0.623$\\
\noalign{\smallskip}\hline\noalign{\smallskip}
\end{tabular}
	
\vspace{0.5em}	
Table notation and description of acoustic features used to build XGBoost model: MFCC\,--\,26 Mel Frequency Spectral Coefficients (13 means \& 13 standard deviations), AF base\,--\,12 common acoustic (dysphonic) features, AF stats.\,--\,84 acoustic features' statistics, AF\,--\, AF base \& AF stats, ALL\,--\, AF \& MFCC.
\end{table}

DenseNet trained with spectrograms had input shape $(46 \times 25)$ and total of $301$ trainable parameters. Even though this setup was considerably less complex, and regularized with drop-out ($0.3$) and l2~weight decay ($1\textrm{e}-04$), the network tended to over-fit after enough training epochs, which we prevented using early stopping that monitored changes in the validation accuracy. This system yielded F1 score of $0.635$ and $0.531$ on the training and validation sets, respectively. The performance on the testing set was $0.562$ (F1 score $0.514$ for class H and $0.609$ for class P). After refitting on the whole development set, the final F1 score got worse on the dedicated testing set to $0.460$ due to difficulties with classification of healthy voices (F1 score $0.239$ for class H and $0.680$ for class P). With raw input data, the network failed to learn any meaningful features (the size of out training dataset is still too small to provide deep learning algorithm to overcome more conventional approaches).

Hyper-parameter optimization for Isolation Forest trained (10-fold validation) with $96$ speech parameters was done the same way as for XGBoost. The best parameters selected upon performance on the development set were as follows: number of estimators ($200$), contamination ($0.4$), maximum features ($0.3$), maximum samples was set to ``auto''. The system yielded F1 score of $0.576\,(\pm\,0.005)$ on the training set and $0.578\,(\pm\,0.023)$ on the validation set. The final F1 score on the dedicated testing set was $0.610$. The performance details (classification matrix and classification report) can be found in Table\,\ref{tab:isolfconf} and Table\,\ref{tab:isolfreport}. This system showed to be sensitive to the number of input features and the performance raised when we selected just a~subset of them.

\begin{table}[htb!]
\caption{Testing CM for DenseNet (MFCC)}
\label{tab:densenetconf}
	\begin{tabular}{p{2.6cm} l l l}
		\hline\noalign{\smallskip}
			& true H & true P & total predicted \\
		\noalign{\smallskip}\hline\noalign{\smallskip}
		predicted H & \textbf{73} & 44 & 117\\
		predicted P & 47 & \textbf{76} & 123\\
		\noalign{\smallskip}\hline\noalign{\smallskip}
		total true & 120 & 120 & accuracy: \textbf{0.621}\\
		\noalign{\smallskip}\hline
	\end{tabular}
\end{table}

\begin{table}[htb!]
\caption{Testing CR for DenseNet (MFCC)}
\label{tab:densenetreport} 
\begin{tabular}{lllll}
\hline\noalign{\smallskip}
& precision & recall & f1-score & no. samples \\
\noalign{\smallskip}\hline\noalign{\smallskip}
class H & 0.624 & 0.608 & 0.616 & 120\\
class P & 0.618 & 0.633 & 0.626 & 120\\
\noalign{\smallskip}\hline\noalign{\smallskip}
avg. / total & 0.621 & 0.621 & \textbf{0.621} & 240 \\
\noalign{\smallskip}\hline
\end{tabular}
\end{table}

\begin{table}[htb!]
\caption{Testing CM for Isolation Forest}
\label{tab:isolfconf} 
\begin{tabular}{p{2.6cm} l l l}
\hline\noalign{\smallskip}
  & true H & true P & total predicted \\
\noalign{\smallskip}\hline\noalign{\smallskip}
predicted H & \textbf{58} & 30 & 88\\
predicted P & 62 & \textbf{90} & 152\\
\noalign{\smallskip}\hline\noalign{\smallskip}
total true & 120 & 120 & accuracy: \textbf{0.617}\\
\noalign{\smallskip}\hline
\end{tabular}
\end{table}

\begin{table}[htb!]
\caption{Testing CR for Isolation Forest}
\label{tab:isolfreport} 
\begin{tabular}{lllll}
\hline\noalign{\smallskip}
  & precision & recall & f1-score & no. samples \\
\noalign{\smallskip}\hline\noalign{\smallskip}
class H & 0.659 & 0.483 & 0.558 & 120\\
class P & 0.592 & 0.750 & 0.662 & 120\\
\noalign{\smallskip}\hline\noalign{\smallskip}
avg. / total & 0.626 & 0.617 & \textbf{0.610} & 240 \\
\noalign{\smallskip}\hline
\end{tabular}
\end{table}

\section{Conclusions}
\label{conclusions}

In search towards robust voice pathology detection system using acoustic (voice) signals, researchers face a~variety of problems. One of the major problems in this field of science is the limited number of available databases. Moreover, commonly used databases~\cite{eye1994voice,Muhammad2017A,mesallam2017development,godino2010pathological} are very hard to combine because of various distinctions such as: a) the databases are labeled in different languages, b) the databases do not comprise same set of speech tasks, c) there is a~variety of voice pathologies unequally distributed across the databases, etc. For these reasons, researchers have used only a~subset of the databases for their experiments providing results related to those carefully selected subset of data. However, this approach limits the possibilities of creating a~robust voice pathology detector. Therefore, in this work, we have conducted experiments on recordings of sustained phonation of the vowel /a/ produced at normal pitch from 4~different databases trying to eliminate those limitation. To the best of our knowledge, this is the first work that uses such a~large set of data to build mathematical models for computerized, objective voice pathology detection.

We researched 3~distinct classifiers within supervised learning and anomaly detection paradigms. We have explored raw waveforms, spectrograms, MFCC and conventional dysphonic features as input data. All experiments were evaluated by the same criteria on the same dedicated testing set. We observed that XGBoost classifier achieved the best results amongst DenseNet and Isolation Forest classifiers. We also observed that not only XGBoost provided the best performance, it could also handle the feature selection (input: all features) by itself in contrary to Isolation Forest classifier, which showed to be sensitive on the feature selection (input: manually selected subset of features). Overall advantage of using speech features and MFCC with XGBoost was the computation speed that allowed us to use exhaustive randomized cross-validated search to optimize the hyper-parameters, as well as the possibility to sort features by importance. This property is useful for clinical interpretability. Nevertheless, we consider these results exploratory due to the limitations of the databases. Reviewing the performances achieved in scenarios with MFCC as input data we conclude that MFCC alone are not reliable enough for robust voice pathology detection, which was also concluded by Ali et al. in~\cite{ali2017intra}. Regarding the DenseNet, we conclude that in voice pathology detection scenarios with this little training data it is better to use inputs with reduced dimensionality in contrary to raw waveform inputs, or make use of transfer learning or data augmentation.

In this article there are several limitations. Firstly, there are limitations inherited from the databases along with new limitations caused by their combination. For instance, some databases have extremely unequal distribution of healthy and pathological classes, most of the databases have alarming inequalities between the number of samples per pathology type (e.g. many pathologies are present less than 3~times in the database), see Figures~\ref{fig:avpd} (AVPD),~\ref{fig:meei} (MEEI),~\ref{fig:pda} (PDA), and~\ref{fig:svd} (SVD). Most databases have no information about severity of the pathology, nor they have information about manifestation of the pathology in phonation, which means that some of the samples might sound as healthy even though they are labelled as pathological and vice versa. Not to mention that recordings are labelled with more than 1~type of pathology, and in different languages, which makes it especially hard to combine or exclude the samples. Since we used 4~available databases, we utilized only the speech task available in all of them: sustained phonation of the vowel /a/ produced at normal pitch. Secondly, even though we have taken countermeasures to balance the classes with sample weights, we did not conduct our experiments separately on subsets of data for different genders.

Up to this point, most papers focused on voice pathology detection used conventional dysphonic features to quantify the underlying voice pathology. In general, these features are conceptually simple, which on one hand is an advantage as these features are clinically interpretable (i.e. clinicians are able to associate the values of the features with known physiological phenomena inside human body)~\cite{Smekal2015b}, but on the other hand these features are often unable to describe the exact voice pathology under focus in a more complex way, especially in advanced stages of the disease (high level of acoustic noise, irregularity of voice, etc.). In future studies, researchers may consider exploring usage of a~more sophisticated set of acoustic features to complexly and robustly describe the voice and speech production deterioration. For instance, such features have already been successfully applied in the field of non-invasive assessment of Parkinson's disease~\cite{Little2009,Tsanas2010,Smekal2015c}. 

With the previously mentioned facts in mind, we think that recordings of the databases commonly used for automatic voice pathology detection should be consulted with clinicians to evaluate the severity of vocal manifestation of the present pathologies. There are standard metrics, which are used to evaluate the quality of voice that can be used for this purpose~\cite{de1997test,Gerratt1993,Kreiman1993,de1997test,Dejonckere2001}. Addition of such information to the databases could provide researchers with a~unique possibility to build models capable of classification and prediction emphasizing the severity of the exact vocal-manifestation (increased acoustic tremor, roughness, breathiness, etc.) of these pathologies.

We also anticipate that deep learning will play its role in robust voice pathology detection on the assumption that more data will be available, or at least reasonable combination of available databases will be made and limitations of these databases will be partially diminished by data augmentation and other countermeasures. In addition, we presume that use of deep learning methods for novelty detection such as deep autoencoder~\cite{pimentel2014review} for modelling the normophonic voice could be an interesting idea for future investigation with prospect to identify even disordered voices that are sparsely distributed across databases.

In summary, acoustic (voice) signals can nowadays be recorded using a~variety of smart devices and processed remotely using modern cloud technologies. In comparison with the conventional perceptual voice quality examination, computerized acoustic analysis of voice signals can provide clinicians with fast, supportive methodology of objective voice pathology detection, assessment, and monitoring that can be used on everyday basis (see Health 4.0). However, to take advantage of such methodology, robust mathematical models capable of precise voice pathology detection must be introduced. Our work proposes the next step towards this goal using various state-of-the-art machine learning algorithms applied to the largest dataset that have been used for the purpose of automatic voice pathology detection.

\begin{acknowledgements}
This study was funded by the grant of the Czech Ministry of Health 16-30805A (Effects of non-invasive brain stimulation on hypokinetic dysarthria, micrographia, and brain plasticity in patients with Parkinson's disease) and the following projects: SIX (CZ.1.05/2.1.00/03.0072), and LO1401. For the research, infrastructure of the SIX Center was used. The authors (P. Harar, Z. Galaz) of this study also acknowledge the financial support of Erwin Schr\"{o}dinger International Institute for Mathematics and Physics during their stay at the "Systematic approaches to deep learning methods for audio" workshop held from September 11, 2017 to September 15, 2017 in Vienna.\\ 
\end{acknowledgements}

{\small
\noindent
\textbf{Compliance with Ethical Standards}\ \
}

\vspace{1em}

{\small
\noindent
\textbf{Conflict of Interest}\ \ The authors declare that they have no conflict of interest.
}

\bibliographystyle{spmpsci}      
\bibliography{ms}   
\end{document}